\PassOptionsToPackage{unicode}{hyperref}
\PassOptionsToPackage{hyphens}{url}
\PassOptionsToPackage{dvipsnames,svgnames,x11names}{xcolor}
\documentclass[
  12pt,
]{article}

\usepackage{amsmath,amssymb}
\usepackage{setspace}
\usepackage{iftex}
\ifPDFTeX
  \usepackage[T1]{fontenc}
  \usepackage[utf8]{inputenc}
  \usepackage{textcomp} 
\else 
  \usepackage{unicode-math}
  \defaultfontfeatures{Scale=MatchLowercase}
  \defaultfontfeatures[\rmfamily]{Ligatures=TeX,Scale=1}
\fi
\usepackage{lmodern}
\ifPDFTeX\else  
\fi
\IfFileExists{upquote.sty}{\usepackage{upquote}}{}
\IfFileExists{microtype.sty}{
  \usepackage[]{microtype}
  \UseMicrotypeSet[protrusion]{basicmath} 
}{}
\usepackage{xcolor}
\usepackage[margin=1in]{geometry}
\makeatletter
\ifx\paragraph\undefined\else
  \let\oldparagraph\paragraph
  \renewcommand{\paragraph}{
    \@ifstar
      \xxxParagraphStar
      \xxxParagraphNoStar
  }
  \newcommand{\xxxParagraphStar}[1]{\oldparagraph*{#1}\mbox{}}
  \newcommand{\xxxParagraphNoStar}[1]{\oldparagraph{#1}\mbox{}}
\fi
\ifx\subparagraph\undefined\else
  \let\oldsubparagraph\subparagraph
  \renewcommand{\subparagraph}{
    \@ifstar
      \xxxSubParagraphStar
      \xxxSubParagraphNoStar
  }
  \newcommand{\xxxSubParagraphStar}[1]{\oldsubparagraph*{#1}\mbox{}}
  \newcommand{\xxxSubParagraphNoStar}[1]{\oldsubparagraph{#1}\mbox{}}
\fi
\makeatother

\usepackage{longtable,booktabs,array}
\usepackage{calc} 
\usepackage{etoolbox}
\makeatletter
\patchcmd\longtable{\par}{\if@noskipsec\mbox{}\fi\par}{}{}
\makeatother
\IfFileExists{footnotehyper.sty}{\usepackage{footnotehyper}}{\usepackage{footnote}}
\makesavenoteenv{longtable}
\usepackage{graphicx}
\makeatletter
\newsavebox\pandoc@box
\newcommand*\pandocbounded[1]{
  \sbox\pandoc@box{#1}%
  \Gscale@div\@tempa{\textheight}{\dimexpr\ht\pandoc@box+\dp\pandoc@box\relax}%
  \Gscale@div\@tempb{\linewidth}{\wd\pandoc@box}%
  \ifdim\@tempb\p@<\@tempa\p@\let\@tempa\@tempb\fi
  \ifdim\@tempa\p@<\p@\scalebox{\@tempa}{\usebox\pandoc@box}%
  \else\usebox{\pandoc@box}%
  \fi%
}
\def\fps@figure{htbp}
\makeatother

\usepackage{booktabs}
\usepackage{caption}
\usepackage{subcaption}
\usepackage{setspace}
\usepackage{float}
\usepackage{threeparttable}
\usepackage{array}
\usepackage{rotating}
\usepackage{multirow}
\usepackage{pdflscape}
\usepackage{afterpage}
\usepackage{amssymb}
\usepackage{pifont}
\usepackage{adjustbox}
\newcommand{\TL}[1]{\label{tab:#1}}
\makeatletter
\@ifpackageloaded{caption}{}{\usepackage{caption}}
\AtBeginDocument{%
\ifdefined\contentsname
  \renewcommand*\contentsname{Table of contents}
\else
  \newcommand\contentsname{Table of contents}
\fi
\ifdefined\listfigurename
  \renewcommand*\listfigurename{List of Figures}
\else
  \newcommand\listfigurename{List of Figures}
\fi
\ifdefined\listtablename
  \renewcommand*\listtablename{List of Tables}
\else
  \newcommand\listtablename{List of Tables}
\fi
\ifdefined\figurename
  \renewcommand*\figurename{Figure}
\else
  \newcommand\figurename{Figure}
\fi
\ifdefined\tablename
  \renewcommand*\tablename{Table}
\else
  \newcommand\tablename{Table}
\fi
}
\@ifpackageloaded{float}{}{\usepackage{float}}
\floatstyle{ruled}
\@ifundefined{c@chapter}{\newfloat{codelisting}{h}{lop}}{\newfloat{codelisting}{h}{lop}[chapter]}
\floatname{codelisting}{Listing}

\makeatother
\makeatletter
\@ifpackageloaded{caption}{}{\usepackage{caption}}
\@ifpackageloaded{subcaption}{}{\usepackage{subcaption}}
\makeatother

\usepackage[round,authoryear]{natbib}
\usepackage{bookmark}

\IfFileExists{xurl.sty}{\usepackage{xurl}}{} 
\hypersetup{
  pdftitle={Female Nomination and Party Vote Share in US Gubernatorial Elections},
  pdfauthor={Paolo Verme},
  pdfkeywords={candidate gender, gubernatorial elections, female
candidates, difference-in-differences, treatment effect heterogeneity},
  colorlinks=true,
  linkcolor={blue},
  filecolor={Maroon},
  citecolor={Blue},
  urlcolor={Blue},
  pdfcreator={LaTeX via pandoc}}

\title{Female Nomination and Party Vote Share in US Gubernatorial Elections}
\author{Paolo Verme\\[4pt]
  {\normalsize University of Bologna, Bologna, Italy}\\[1pt]
  {\normalsize \texttt{paolo.verme@gmail.com}}}
\date{August 2026}

\begin{document}

\maketitle
\begin{abstract}
What happens to a party's vote share when it nominates a woman for executive office? The evidence on female candidates is dominated by legislative races and by designs that pool parties and periods into a single average effect, which is typically null. This paper builds a new county-level panel of US gubernatorial elections — 578 races in 49 states from 1980 to 2024, reconstructed and audited from primary sources — and estimates party-specific, time-varying nomination effects with a stacked difference-in-differences design around within-state transitions from male to female nominees. The average null conceals a sharp asymmetry. Republican transitions to a female nominee since 2016 are followed by a county-level vote-share decline of 6.5 percentage points, robust across cohort, weighting and inference checks. No comparable average differential is detected for Democratic nominees. Survey-ballot diagnostics are compatible with demand-side discrimination but cannot identify that channel.
\end{abstract}

\setstretch{1.5}
\vspace{6pt}

\noindent\textbf{Keywords:} candidate gender; gubernatorial elections; female
candidates; difference-in-differences; treatment effect heterogeneity

\noindent\textbf{JEL classification:} D72; J16; C23

\newpage

\section{Introduction}\label{sec:intro}

The question of whether the US electorate is willing to elect a woman to
the presidency has moved from academic debate into public political
discourse. After the 2024 presidential election, Michelle Obama argued
that, ``sadly, we ain't ready,'' adding, ``You're not ready for a
woman'' \citep{obama2025look}. Kamala Harris raised a similar concern. During her 2020 presidential campaign she
described electability as ``the elephant in the room about my
campaign'' \citep{harris2019axios}. In her account of selecting a running
mate in 2024, she wrote that the ticket was already ``asking a lot of
America'' because voters were being asked to accept a woman \citep{harris2025memoir}. These statements do not establish voter
discrimination, but they pose a clear empirical question: what happens
to a party's electoral performance when it nominates a woman for
executive office?

Presidential elections cannot answer that question with any statistical
power. Only two major-party presidential nominees in US history have
been women, both were Democrats, both ran in recent elections, and both
lost to the same Republican candidate. The historical record is therefore
politically salient but empirically too thin to separate candidate
gender from party, period, opponent and election-specific conditions.
Gubernatorial elections provide a broader and institutionally relevant
setting. They are executive elections, like presidential elections, but
they recur across states and election cycles and generate repeated
within-party transitions from male to female nominees.

Women remain markedly underrepresented in this setting as well. Between 1980 and 2024, women accounted for 151 of 1,156 major-party
gubernatorial nominations, or 13.1 percent. Female nominees are more
common among Democrats with 103 of 578 Democratic nominees being women,
compared with 48 of 578 Republican nominees, corresponding to 17.8 and
8.3 percent, respectively. Before asking why this imbalance persists, a
prior empirical question is what happens to a party's vote share when it
nominates a woman. This paper addresses this question using a newly
constructed panel of US gubernatorial elections.

This question is not equivalent to a direct test of voter discrimination.
Nominee gender is chosen jointly with candidate experience, ideology,
fundraising, primary competition and opponent quality. The estimand in
this paper is therefore the reduced-form electoral-performance
differential associated with a party nominating a woman for governor. That differential may reflect voter responses, candidate and
party selection, or both. Separating these channels is a central
limitation of observational election data and guides the interpretation of results throughout the paper.

The existing evidence on this question is mixed. Experimental and quasi-experimental studies typically
find small or null average effects, whereas observational studies
occasionally document disadvantages for female candidates in executive
races. Most existing studies also estimate a single average effect over
comparatively short windows, so they cannot speak to how the relationship
has changed over time.

This paper addresses this gap by building a new panel data set disaggregated at the county level for US
gubernatorial elections in 49 states from 1980 to 2024. The empirical strategy combines
two-way fixed effects, used as a full-panel descriptive benchmark, with a
stacked difference-in-differences design around recurrent
\(0\rightarrow1\) transitions to a female nominee. Because nomination is
assigned at the state-year level, the stacked estimates are reported both
on the 578-race state-year panel and on the county-level panel, which preserves
local electoral outcomes and partisan baselines. The state-year estimates assess
sensitivity to aggregation and county weighting.

The main finding is an asymmetry in female-nomination performance differentials across parties. For Democratic nominees, the full-panel and recent-cohort stacked estimates are indistinguishable from zero. For Republican nominees, the reported estimates are consistently negative, although their precision depends on the level of aggregation. The post-2016 county-level stacked estimate is \(-6.5\) percentage points, and every cohort-specific estimate is negative; the corresponding state-year estimate is \(-3.5\) points but imprecisely estimated. Evidence of a negative Republican pre-trend limits causal interpretation. The result is therefore best understood as a robust reduced-form regularity rather than as an estimate of voter discrimination.

The paper proceeds as follows. Section~\ref{sec:litreview} reviews the literature. Section~\ref{sec:strategy} describes the empirical
strategy explaining how the data constraints and identification issues have been addressed. Section~\ref{sec:results} presents the empirical results, including TWFE estimates, stacked DiD estimates, and pre-trend diagnostics. Section~\ref{sec:robustness} covers a wide range of robustness checks and Section~\ref{sec:conclusion} concludes. Tables and figures follow the references. The replication package that accompanies the submission carries the panel, the build and analysis code, and the full set of tables behind the figures and the numbers quoted in the text.

\section{Literature}\label{sec:litreview}

Whether female candidates face electoral disadvantages has been examined
extensively in economics and political science, and the evidence remains
largely inconclusive.

Political science evidence initially suggested little systematic
disadvantage for women. In congressional elections, female candidates perform similarly to male candidates conditional on party and incumbency \citep{darcy1987, palmer2008, dolan2014, dolan2004}. Where candidate sex does move votes it tends to do so conditionally rather than on average, through the stereotypes voters bring to a race and the information they seek about it \citep{sanbonmatsu2002, ditonto2014, ono2019}, and through how candidates campaign on gender \citep{herrnson2003}. A
meta-analysis of sixty-seven factorial survey experiments similarly finds
little evidence of an average penalty in stated vote choice. If anything,
respondents slightly favor female candidates, by roughly two percentage
points, though with pronounced heterogeneity across partisan contexts
\citep{schwarz2022}. Complementary experimental work identifies mechanisms
that can nonetheless disadvantage women in practice: voters hold female
candidates to more stringent qualification standards \citep{bauer2020} and
withhold support for female candidates out of concern that other voters
will not elect a woman \citep{corbett2022}.

Theory and evidence suggest that these average null effects may mask
important heterogeneity.
Role congruity theory predicts stronger gender bias in executive
positions, where leadership traits are more closely associated with
masculinity \citep{eagly2002}. Consistent with this, \citet{anzia2022}
find that women running in US local elections outperform men in city
council and school board races but face a substantial disadvantage in
mayoral contests, the one executive office in their data, and
\citet{fulton2009} shows that apparent gender parity in congressional
races masks a penalty that is offset by the higher average quality of
female candidates. Legislative elections are more party-centered whereas executive elections weigh candidate-specific traits more heavily.

The economics literature, focused on causal identification through
quasi-experimental designs, reaches similar conclusions. Studies
exploiting close elections and natural experiments find little evidence
of systematic gender penalties in US races. \citet{ferreira2014} find
that electing a female mayor has no detectable effect on policy
outcomes, with female mayors if anything enjoying a larger incumbent
advantage. \citet{anastasopoulos2016} finds no gender penalty in U.S.
House general elections when female nominees emerge from close
primaries, and \citet{broockman2016}, studying quasi-random ballot
positions in Illinois Republican presidential primaries, find
substantial discrimination against nonwhite candidates but essentially
none against women.

The European evidence sharpens the distinction between voter response and party gatekeeping that this paper's estimand combines. On voter evaluation, \citet{lebarbanchon2022} document a vote-share penalty for female candidates in mixed-gender run-offs in French parliamentary elections, concentrated where gender attitudes are more traditional, together with strategic nomination: parties reserve male candidates for the more contestable districts. \citet{esteve2012} separate the two margins directly in Spanish Senate elections, where open lists let voters deviate from party rankings: female candidates attract, if anything, more votes than comparable men, and women's disadvantage arises from parties assigning them to worse list positions.

On party selection and candidate quality, the Italian and Swedish quota literatures show that the binding constraint is often the party, not the voter: gender quotas in Italian municipal elections raised the average quality of elected politicians \citep{baltrunaite2014}, and municipalities briefly exposed to Italy's 1993--1995 quota kept electing more women after it was struck down, consistent with quotas eroding stereotypes rather than forcing unpopular candidates on voters \citep{depaola2010}; in Sweden, quotas raised the competence of the political class by displacing mediocre men \citep{besley2017}.

On access to executive leadership --- the margin closest to this paper's setting --- the barrier appears steepest at the top: promotion discrimination within Swedish politics intensifies toward leading positions \citep{folke2016}, quotas in Spain and Italy lift women into councils but not into list leadership or mayoralties \citep{bagues2021, spaziani2022}, and in Germany the quasi-random election of a female mayor raises the subsequent success of female council candidates, indicating that exposure to women in executive office itself shifts outcomes \citep{baskaran2018}. Most of this evidence comes from proportional or list systems and from municipal government, where party gatekeepers mediate between voters and candidates. The US gubernatorial setting is the candidate-centered complement: a two-party, statewide executive race in which voters directly choose among individually named nominees for statewide executive office rather than among party lists, so the electoral consequence of nominating a woman is observed directly rather than through list placement.

Supply-side explanations are central to interpreting these
results. Women are less likely to run for office due to differences in political ambition, recruitment, and social constraints \citep{fox2014, fox2004, fox2010, lawless2012, teele2018, bernhard2021}, implying that female candidates may be positively selected on quality. The gap is partly strategic rather than dispositional. Women weigh their prospects before entering and are less likely to seek higher office when they expect to lose \citep{fulton2014}, and the states and offices where they run differ systematically \citep{sanbonmatsu2006}. Because incumbents rarely lose, the timing of open seats shapes when women can enter at all \citep{schwindt2005, gaddie2000, windett2011}. There is also evidence of a ``quality gap''. Women who run are, on average, more qualified than their male counterparts \citep{fulton2009}. A related literature shows that female leaders shift policy toward the priorities of women \citep{chattopadhyay2004}, govern with less corruption and patronage \citep{brollo2016}, and can be brought into office without lowering the quality of the political class \citep{besley2017}; that they alter the substance of representation and deliberation once in office \citep{anzia2011b, osborn2012, karpowitz2014}; and that their presence affects women's engagement with politics more broadly \citep{atkeson2003, wolbrecht2007, pande2012}.
Together, these findings suggest that observed electoral outcomes
reflect an equilibrium between voter preferences and candidate selection rather than a direct measure of voter bias.

Recent research suggests that the relationship may also be time-varying. Studies of U.S. elections show that
sexism was a powerful predictor of vote choice in 2016
\citep{schaffner2018, winter2023}, while
partisan coalitions have realigned along educational and cultural dimensions
\citep{sides2018, mutz2022} and partisan affect has hardened \citep{iyengar2019}, potentially
increasing the salience of candidate characteristics such as gender in polarized
environments. Earlier cycles show that a favorable national climate can produce
a burst of female candidacies without a durable shift in representation
\citep{cook1998}.

Least settled, and closest to this paper's finding, is whether any of this
operates symmetrically across parties. Americans associate the Republican party
with masculine traits and the Democratic party with feminine ones
\citep{winter2010}, and voters use candidate sex as an ideological cue,
perceiving female candidates as more liberal than comparable men
\citep{koch2002}. Experimental work manipulating the sex of a
\textit{Republican} candidate specifically finds support falling among some
co-partisans \citep{king1997}, and the pool of Republican women who run is
shaped by whether they see themselves as fitting an increasingly conservative
party \citep{thomsen2015}. Together these suggest the demand and supply sides
may both operate differently on the two sides of the aisle, the asymmetry this
paper documents in vote shares.

Taken together, existing studies leave unresolved whether the electoral
differential associated with nominating a woman varies jointly by office
type, party, and political period. The 45-year gubernatorial panel
addresses that gap in repeated executive contests while treating observed
vote-share differences as equilibrium outcomes of voter response and
party selection.

\section{Empirical Strategy}\label{sec:strategy}

\subsection{The Core Identification Issues and Their
Treatment}\label{sec:identification}

Two constraints shape the design. The first is statistical power. Estimating party-specific female-nomination effects requires enough state-years in which a party switches from a male to a female nominee, and enough contemporaneous states that do not, to form credible comparison groups. This is demanding in gubernatorial elections because female major-party nominees remain historically rare. The paper addresses this constraint by rebuilding a long state-year and county-level panel of US gubernatorial elections from 1980 to 2024, which provides the longitudinal and spatial coverage needed to study nomination transitions separately by party.

The second constraint is endogenous nomination. Candidate gender is not assigned at random. Parties choose nominees jointly with candidate experience, ideology, fundraising capacity, primary competition, opponent quality and the electoral environment. If parties are more likely to nominate women in difficult races, an observed vote-share penalty may reflect the conditions under which women are nominated rather than voter bias against female candidates. Conversely, if women who reach the ballot are positively selected on quality, observed estimates may understate any voter penalty. The empirical problem is therefore not simply whether voters respond to candidate gender, but how to interpret the electoral performance associated with actual nominations.

A simple equilibrium framework clarifies the issue. Parties select candidates characterized by quality \(q\) and gender \(g \in \{\text{male}, \text{female}\}\), while voters choose between nominees on the basis of party affiliation, candidate attributes and social identity considerations. Let \(\theta\) denote the political environment, including polarization, the importance of identity and partisan sorting. Voters may attach a context-dependent weight \(\beta(\theta)\) to candidate gender, and parties anticipate those responses when selecting nominees. In equilibrium, the observed relationship between nominee gender and vote share therefore combines voter response and endogenous candidate selection. This logic is consistent with models of political selection and citizen-candidate entry \citep{besley1997, caselli2004}, in which candidate pools depend on both incentives and constraints.

The available gubernatorial data cannot fully separate these channels. The panel observes election returns, party, incumbency, open-seat status and selected candidate characteristics, but it does not observe party recruitment strategies, candidate quality as perceived by voters, primary-field composition, fundraising expectations, or the counterfactual male nominee a party would otherwise have selected. The estimates should therefore be read as reduced-form female-nomination performance differentials: the electoral consequences associated with a party nominating a woman for governor. The analysis probes this interpretation through controls for lagged vote share and incumbency, stacked comparisons around nomination transitions, party-specific estimates, candidate-experience adjustments, aggregation checks, cohort-level robustness, and survey-ballot diagnostics.

The empirical strategy addresses these challenges through a combination of data construction,
empirical design, and robustness checks. The outcome is the percentage of
votes received by each party's candidate, measured at county level, across
all gubernatorial elections from 1980 to 2024. This long panel allows us to leverage
within-state variation in candidate gender over time while controlling
for state and year fixed effects, mitigating confounding factors from
time-invariant state characteristics and common shocks. The analysis then uses a
stacked DiD design around recurrent \(0\rightarrow1\) nomination
transitions. The TWFE full-panel estimate is a
descriptive benchmark. The recent stacked design uses clean
two-election comparison windows and is interpreted under a parallel-
trends assumption; it is reported on the 578-race state-year panel, the
level at which nomination is assigned, with the county-level counterpart alongside it; Section \ref{sec:robustness} reports
what separates them. Candidate-experience checks, aggregation diagnostics
and survey evidence probe possible explanations but do not convert the
nomination differential into a voter-discrimination parameter.

\subsection{Data}\label{sec:data}

The empirical analysis uses a newly constructed county-level panel of US gubernatorial general elections from 1980 to 2024. The final file contains 35,135 county-election observations covering 578 state-year races in 49 states. It includes the regular gubernatorial cycle and the off-cycle elections in New Jersey, Virginia, Kentucky, Louisiana and Mississippi. Alaska is excluded because it reports by borough rather than by county and is not consistently covered in the historical county-return sources. The District of Columbia is not covered because it has no gubernatorial office. Of the 35,135 county observations, 31,929 carry a lagged party vote share and enter specifications that condition on prior partisan performance. Table \ref{tab:summary} reports summary statistics.

The panel improves on existing data sources in coverage, coding and auditability. The historical backbone is the county-level gubernatorial return file of \citet{algara2021}, used here through 2020. Keeping that source through 2020 rather than switching earlier to precinct-archive conversions avoids several recent-party-coding problems and restores county rows missing from the pre-existing corrected panel, including substantial coverage gains in Indiana in 2016 and 2020, Washington in 2016 and New York in 2018. Elections after 2020 are then assembled from official precinct archives, state canvasses, Secretary of State PDFs and certified web returns. This produces a unified source spine rather than a splice of partly corrected intermediate files. Odd-year elections after 2020, which are absent from the standard even-year precinct archives, are filled separately from certified sources. Gaps in the 2022 archive are also filled where possible, including Tennessee and New York, where the archive either omits the race or loses fusion-ballot information.

The rebuild matters because several errors in available files are concentrated in precisely the cells that matter for this paper. First, candidate gender is re-coded from the CAWP historical records and then audited against the nominee names carried in the returns themselves. The pre-audit gender file contained 122 female-nominee state-years. The name audit adds 29 further female major-party nominations, yielding 151 female nominations in the final panel: 103 Democratic and 48 Republican. Ten races have women on both major-party tickets, so 141 of the 578 races include at least one female major-party nominee. This correction is consequential because the identifying variation is sparse and party-specific.

Second, the vote-return build explicitly repairs party-label and denominator problems in the recent precinct archives. The conversion maps fusion-party labels to the appropriate major party, excludes pseudo-candidates such as blanks and undervotes from candidate totals, avoids double-counting where total-mode and vote-mode records coexist, and treats zero major-party shares as hard errors rather than valid observations. A separate correction layer applies certified county-level shares where the candidate votes are correct but the total-vote denominator is inflated by ballot lines that the precinct converter does not recognize. In the final build, 43 county rows require this correction. These corrections were essential. The most serious affected state-years were female-Democratic-nominee races in the recent window.

Third, the panel revises controls that are central to the identification strategy. Incumbency and open-seat status are taken directly from the historical source where available and derived for recent elections using a prior-winner rule with manually recorded succession exceptions. This fixes cases in which succession governors who took office mid-term and then ran as incumbents were coded as open-seat candidates. Because incumbency and open seats enter both the main specifications and the selection diagnostics, these corrections are not cosmetic. Candidate-experience variables are also audited against the nominee names in the returns, which prevents experience records from attaching to the wrong person.

The build is auditable at each step. County returns for 1980--2020 come from the \citet{algara2021} archive (32,045 observations, 527 state-years); 2022 and 2024 come from the MIT Election Data and Science Lab precinct archives; New Jersey 2021, Virginia 2021, Tennessee 2022, New York 2022, Louisiana 2023, Mississippi 2023 and Kentucky 2023 are filled from certified state sources. Where the historical archive and the precinct archives overlap, in 2016--2020, 99.6\% of county observations agree to within one hundredth of a percentage point, and the archive restores 143 county observations missing from the earlier panel. The correction layer carried by the previous panel contained 171 county rows; rebuilding from primary sources makes 128 of them redundant, and the 43 that remain are the denominator corrections described above. For 2021--2024, where incumbency has to be derived, the prior-winner rule agrees with the archive's own seat-status field on 98.1\% of Democratic and 96.4\% of Republican cases in the years where both exist; its known failure mode is mid-term succession, which is why succession governors are enumerated by hand.\footnote{The seven state-years previously coded as open seats despite a sitting governor running are Kate Brown (Oregon 2018), Kathy Hochul (New York 2022), Daniel McKee (Rhode Island 2022), Kim Reynolds (Iowa 2018), Henry McMaster (South Carolina 2018), Mike Parson (Missouri 2020) and Kay Ivey (Alabama 2018). Alabama 2018 is the only case the historical archive itself miscodes, and it is consequential: it was the single positive Republican cohort estimate before the correction.} Nominee names are checked automatically against the returns, and all 151 female-nominee rows match on surname.

Four limitations remain. Minnesota 1998 is dropped because the historical archive records zero votes for the Democratic-Farmer-Labor nominee in every county; both nominees were men, so this affects controls and lags rather than treated observations. Arizona and California 2022 are carried forward from the earlier checked panel, California with two-party shares only. Three state-years remain short of their full county complement (Indiana 2024, Idaho 2022 and Arizona 1982). Illinois 1986 and Louisiana's jungle-primary years are retained under the source's party-line conventions. None of these is consequential for the estimates. The audit matters because the defects it corrected were not randomly distributed: zeroed returns, inflated denominators and mis-assigned incumbencies were concentrated in races with a female nominee and in the most recent decade, which are precisely the cells that identify the estimates.

The dependent variables are party vote shares. The Democratic outcome is the Democratic share of all ballots cast, while the Republican outcome is defined analogously. The panel also retains two-party vote shares for transparency and for constructing lagged outcomes. Lagged county-level vote share is the party's share in the previous gubernatorial election in the same county and is coded as missing when consecutive elections are more than six years apart. Nominee gender, and hence treatment status, varies at the state-year level: all counties within a state share the same Democratic and Republican nominees.

Finally, the survey diagnostics use the CCES cumulative file for 2006--2024. The gubernatorial election calendar for those diagnostics is derived from the rebuilt returns rather than hard-coded, and the county-level CCES panel is aggregated using respondent county identifiers where available, with state-year aggregates used as fallback. The CCES data are used only for the survey-ballot and individual-level loyalty diagnostics; they do not enter the main vote-share estimates. The seven build scripts that produce the panel from the raw sources are part of the replication package.

\begin{center}
\textit{[Table \ref{tab:summary} approximately here]}
\end{center}

\subsection{Models}\label{sec:models}

The two-way fixed effects (TWFE) estimator is the natural starting point
for this setting. State fixed effects absorb time-invariant cross-state
heterogeneity (differences in partisan baseline, political culture, and
the historical propensity to nominate women), and year fixed effects
absorb shocks common to all states in a cycle, such as the
presidential-year and midterm swings. The TWFE \(\beta\) coefficient is
thus identified from states that switch nominee gender in a given cycle
relative to states that do not. 

The model further conditions on the
lagged county-level vote share and incumbency status to absorb
pre-existing partisan trends and the well-documented incumbency
advantage. This design cannot eliminate endogenous selection of female
candidates into favorable or unfavorable electoral environments, but it
substantially reduces confounding relative to cross-sectional or pooled
OLS comparisons. The core TWFE model is as follows: 

\begin{equation}
\text{dem\_share}_{ist} = \alpha_s + \gamma_t + \beta \cdot \text{FDN}_{st} + \delta \cdot \text{dem\_share}_{ist-1}^{\text{2p}} + \theta \cdot \text{Incumbent}_{st} + \varepsilon_{ist}
\label{eq:twfe}
\end{equation}

The coefficient \(\beta\) measures the average within-state change in the Democratic vote share associated with nominating a female Democratic candidate, relative to states and years without such a nomination, conditional on the lagged vote share and incumbency.

The dependent variable in equation \ref{eq:twfe} is the Democratic share
of all ballots cast in county \(i\), state \(s\), and election year \(t\):
\[
\text{dem\_share}_{ist}
= 100 \times
\frac{\text{Democratic votes}_{ist}}{\text{total votes}_{ist}}.
\]
The treatment indicator is
\[
\text{FDN}_{st} =
\begin{cases}
1, & \text{if the Democratic gubernatorial nominee in state } s \text{ in year } t \text{ is female},\\
0, & \text{otherwise.}
\end{cases}
\]

In equation \ref{eq:twfe}, \(\alpha_s\) are state fixed effects, \(\gamma_t\) are year fixed
effects, and standard errors are clustered by state. The subscript
\(ist\) indexes county \(i\) in state \(s\) at election year \(t\). The
superscript \(\text{2p}\) on the lagged vote share denotes the two-party
share (with only the Democratic and Republican votes in the
denominator), whereas the outcome \(\text{dem\_share}_{ist}\) uses total
votes including third-party ballots. The two-party measure provides a
cleaner baseline for the prior partisan environment by removing noise
from third-party candidacies. The symmetric Republican model
replaces the outcome with \(\text{rep\_share}_{ist}\), the treatment
with \(\text{FRN}_{st}\), and conditions on the lagged Republican vote
share and a Republican incumbent indicator.

TWFE suffers from ``forbidden'' comparisons \citep{goodman2021, baker2022}. TWFE implicitly uses
already-treated units as controls for units that switch later. Here, a
state nominating a woman in one cycle may enter comparisons around
another state's transition, even though nomination status can
subsequently reverse. If the performance differential varies across
events, TWFE takes a weighted average of all such pairwise comparisons,
and the weights can be negative for early-adopting cohorts, causing the
pooled estimate to be a misleading mixture of cohort-specific effects
and even to reverse sign relative to every cohort-specific effect
\citep{goodman2021}.

This concern is addressed by constructing explicit two-election
blocks around recurrent \(0\rightarrow1\) transitions, using the stacked
DiD of \citet{baker2022}, which builds a separate comparison group for each
cohort. Let \(\mathcal{C}\) denote the set of switch cohorts included in a
particular stacked specification. In different tables, \(\mathcal{C}\) may
contain all estimable transitions, only pre-2016 transitions, or only
2016--2024 transitions. The matching rule is unchanged across these variants:
a control state must hold gubernatorial elections in \textit{the same two
calendar years} as the treated state and must have no female nominee of the
treated party in either. For a cohort \(c \in \mathcal{C}\) (a state
transitioning from FDN\(=0\) to FDN\(=1\)), the pooled stacked equation is:

\begin{equation}
\text{dem\_share}_{istc} = \alpha_{s \times c} + \gamma_{t \times c} + \beta \cdot \text{FDN}_{stc} + \theta \cdot \text{Incumbent}_{stc} + \varepsilon_{istc}
\label{eq:stacked}
\end{equation}

The coefficient \(\beta\) measures the average stacked difference-in-differences change in the Democratic vote share after a Democratic nomination switches from male to female, relative to clean control states in the same cohort block, conditional on incumbency.

The dependent variable in equation \ref{eq:stacked} is again the Democratic
share of all ballots cast, now indexed by cohort block \(c\):
\[
\text{dem\_share}_{istc}
= 100 \times
\frac{\text{Democratic votes}_{ist}}{\text{total votes}_{ist}}.
\]
The treatment indicator is
\[
\text{FDN}_{stc} =
\begin{cases}
1, & \text{if the Democratic nominee in state } s \text{ and year } t \text{ is female},\\
0, & \text{otherwise.}
\end{cases}
\]
The Republican rows of Table \ref{tab:stacked_windows} use the symmetric
replacement: \(\text{rep\_share}_{istc} = 100 \times
\text{Republican votes}_{ist}/\text{total votes}_{ist}\) and
\(\text{FRN}_{stc}\).

In equation \ref{eq:stacked}, \(\alpha_{s \times c}\) and \(\gamma_{t \times c}\)
are cohort-specific state and year fixed effects. Note that the lagged
vote share is omitted. Each cohort is a two-period panel, so the
pre-switch outcome is the implicit baseline absorbed by the
cohort-specific fixed effects. Cohort-specific diagnostic estimates
replace \(\beta\) with \(\beta_c\); the main pooled tables impose the
common \(\beta\) reported in equation \ref{eq:stacked}. Some blocks are
degenerate and some cohorts may have no clean control state at all and
drop out entirely, so each table reports the number of usable cohorts it
actually estimates.

Although outcomes are observed by county, treatment varies at the state-year
level, so counties do not constitute independent treatment assignments. The
county panel permits adjustment for local lagged partisan support in the TWFE
specifications and captures within-state variation in electoral responses in
both designs, but it may implicitly give greater weight to states containing
more counties. Tables \ref{tab:temporal} and \ref{tab:stacked_windows} therefore
report every estimate on the 578-race state-year panel alongside its county
counterpart, and Section \ref{sec:robustness} reports the wild-cluster inference
that accompanies them.

The stacked design is one of several estimators proposed for staggered adoption with
heterogeneous effects. \citet{callaway2021} estimate group-time average treatment
effects $ATT(g,t)$ against a never-treated or not-yet-treated comparison group and then
aggregate; \citet{sun2021} saturate an event study in cohort-by-relative-period
interactions and reweight; \citet{borusyak2024} impute untreated potential outcomes from
the never-treated cells; and \citet{dechaisemartin2020} restrict attention to
comparisons with a determinate sign. All four share the same diagnosis and differ in
which clean comparisons they retain and how they average them. The paper estimates the stacked
design in the main text because it makes the comparison group explicit --- each cohort's
counterfactual is a named set of states untreated in both periods. Robustness checks below
then ask whether the result changes when the same identifying variation is aggregated in
alternative ways.

\section{Results}\label{sec:results}

Table \ref{tab:main} reports TWFE estimates building up the
specification one control at a time. For FDN, the
point estimate is never significant. It shifts from \(-2.03\) pp (no
controls) to \(+0.32\) pp (\(SE = 1.31\), \(p = 0.809\)) once the lagged
vote share and incumbency are included. Both controls are significant,
raising \(R^2\) and moving the FDN coefficient toward zero. For FRN
the estimate is negative in every specification and, with the complete
specification, statistically significant \(-4.07\) pp (\(SE = 1.54\), \(p = 0.011\)), and remarkably stable across
control sets.

\begin{center}
\textit{[Table \ref{tab:main} approximately here]}
\end{center}

Table \ref{tab:temporal} investigates temporal heterogeneity using the full 1980--2024 panel, the pre-2016 period, and the recent 2016--2024 period.
Each row re-estimates the preferred TWFE specification from column (4) of Table
\ref{tab:main}, and reports both the county estimate and the corresponding
state-year estimate at the level where nomination is assigned.

For Democratic nominees, the TWFE estimates are null in all three periods and at
both levels of aggregation. The full-period county estimate is +0.32 pp
(\(SE=1.31\)), the pre-2016 estimate is +0.91 pp (\(SE=1.82\)),
and the recent estimate is +1.91 pp (\(SE=2.37\)). Collapsing to the
state-year level gives similarly null estimates: -1.31, -0.90 and +2.68 pp.
Thus the Democratic side remains a descriptive null for all specifications with sign that changes both across periods and across the county and state panels.

For Republican nominees, TWFE estimates are negative in all
three periods and at both levels of aggregation. The county estimates are all large and statistically significant: -4.07 pp in the full panel, -4.67 pp before 2016, and -7.86 pp in
2016--2024. The corresponding state-year estimates are consistently negative but smaller and imprecise: -1.59, -2.01 and -1.69 pp, respectively for the three periods and all non-significant. This contrast is useful because it separates the county-weighted TWFE
benchmark from the cleaner stacked comparison in Table \ref{tab:stacked_windows},
where the recent Republican estimate remains negative at the state-year level.

\begin{center}
\textit{[Table \ref{tab:temporal} approximately here]}
\end{center}

The stacked DiD draws each transition's counterfactual from states
untreated in both the pre- and post-switch windows, removing cross-event
comparisons and estimating a female-nomination performance differential event by
event. Table \ref{tab:stacked_windows} applies the same temporal split used for the TWFE estimates. For comparability with Table
\ref{tab:temporal}, all estimates in Table \ref{tab:stacked_windows} include
own-party incumbency whereas the lagged vote share remains omitted because, in a
two-period stack, the pre-period outcome is the baseline.

The Democratic stacked estimates are indistinguishable from zero in all three
cohort sets and at both levels of aggregation. As for TWFE estimates, signs change across periods and level of aggregation. For Republican nominees, the
county estimates are negative in all three cohort sets and statistically
significant in the full period (-5.26 pp, \(SE=2.24\)) and recent period
(-6.50 pp, \(SE=1.90\)). The state-year estimates are also negative in all
three cohort sets but imprecise: -3.32 pp in the full period, -3.22 pp before
2016, and -3.47 pp in the recent cohort set. Thus the incumbency-adjusted
stacked table aligns the sign of the Republican estimate across periods, but it
does not by itself establish a significant state-year penalty. 

\begin{center}
\textit{[Table \ref{tab:stacked_windows} approximately here]}
\end{center}

These findings point to a cross-party asymmetry which can be formally tested. Table
\ref{tab:asymmetry} stacks both parties' blocks into one panel for the same
full-period, pre-2016 and recent cohort sets used in Table \ref{tab:stacked_windows} ---
own-party share as outcome, own-party nomination as treatment --- and
interacts treatment with a Republican indicator, controlling for own-party incumbency.
That interaction is the difference between the parties. 

Overall, the Republican--Democratic differentials are negative in all specifications and
fall within a relatively narrow range, from \(-4.89\) to \(-6.94\) pp. At the county
level, the differentials range from \(-6.50\) to \(-6.94\), with statistically
significant estimates in the full-period and recent-cohort specifications. At the
state-year level, the differentials remain negative, ranging from \(-4.89\) to
\(-5.46\), but are not statistically significant. These
results reinforce the county-level asymmetry finding and the inconclusive state-year
evidence reported for the TWFE and stacked DiD estimates.

The weaker state-year evidence is consistent with limited statistical power and may help
explain why previous studies rarely found a significant relationship between female
nominations and electoral outcomes. It underscores the importance of observing a
sufficiently large number of mixed-gender races and of using the county panel in a context
where female gubernatorial nominees remain scarce. Wild-cluster bootstrap inference does
not overturn the county-level Republican results: the bootstrap \(p\)-value is 0.018 for
the full-panel TWFE estimate and 0.002 for the recent stacked estimate, against 0.091
for the state-year stacked estimate. The county panel does not create additional treatment variation, but it
preserves local electoral responses and lagged partisan baselines within each statewide
race, improving precision while keeping inference clustered at the state level.

\begin{center}
\textit{[Table \ref{tab:asymmetry} approximately here]}
\end{center}

Table \ref{tab:pretrend} reports the pre-trend diagnostic, which consists of the post-2016 switch cohorts estimated in the same specification used for the recent stacked estimate. The Democratic pre-period coefficients are small and jointly insignificant. For Republicans, the \(k=-2\) coefficient is \(-4.12\) pp (\(SE = 1.98\)), significant at the five percent level, although the joint test over both pre-periods does not reject (\(p = 0.114\)). Parallel trends therefore cannot be confirmed for the Republican estimate. This is a limitation of the available identifying variation. Even after rebuilding the panel, the pre-trend diagnostic rests on only 13 Republican
switch events. The diagnostic is therefore informative but not conclusive.

\begin{center}
\textit{[Table \ref{tab:pretrend} approximately here]}
\end{center}

\section{Robustness Checks}\label{sec:robustness}

This section provides a battery of robustness checks that ask whether the results depend on how the stacked design is built, on thin identifying variation, on selection into female nomination, on observable candidate quality, on survey interpretation, or on limited power.

\begin{figure}[H]
\centering
\includegraphics[width=\textwidth]{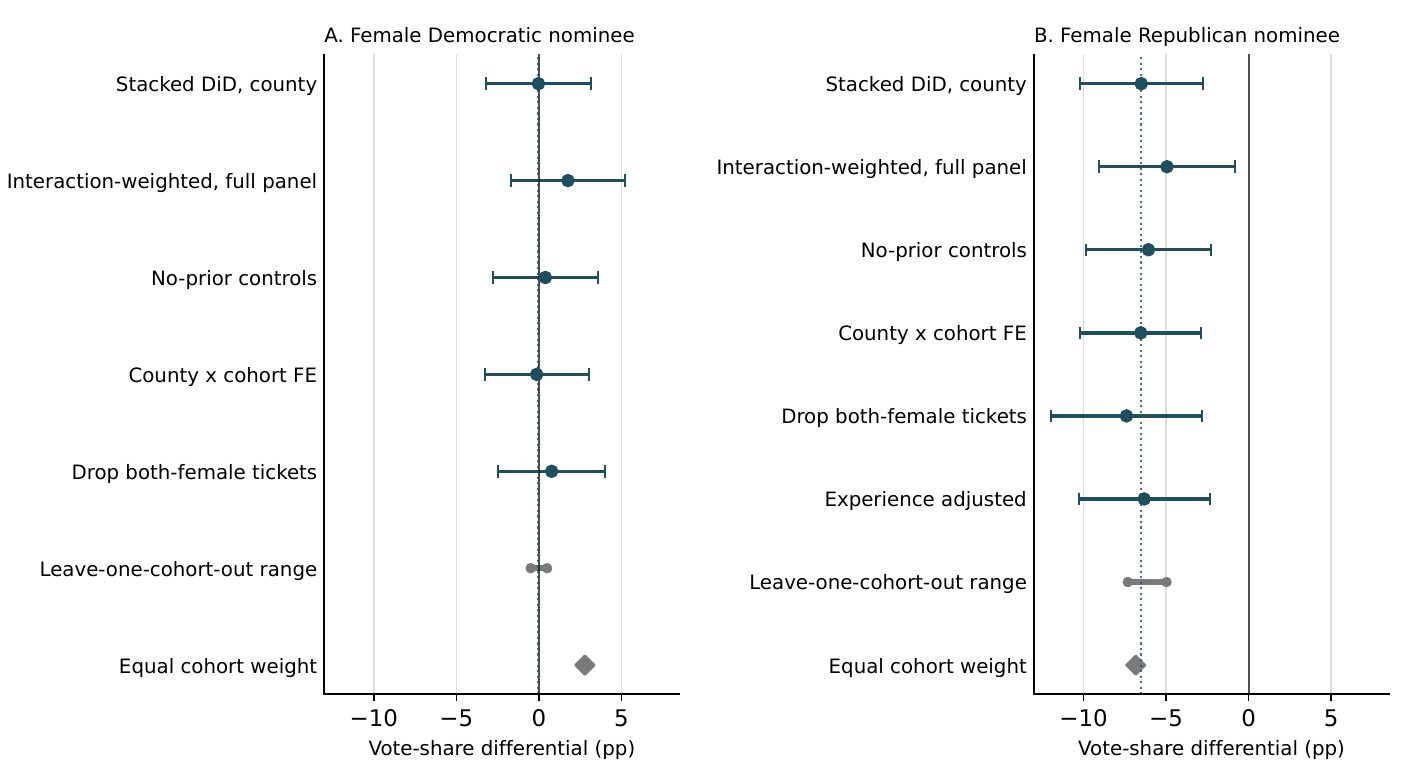}
\caption{Robustness of the Recent Stacked Estimates to How the Design Is Built}\label{fig:robustness}
\end{figure}

\textit{How the design is built.} Figure \ref{fig:robustness} examines whether the county-level stacked estimates depend on the construction of the comparison group, the treatment of heterogeneity across cohorts, observable candidate characteristics, or particular nomination switches. The county estimates in Table \ref{tab:stacked_windows} are the baselines, marked by the dotted vertical lines. The interaction-weighted row is a full-panel benchmark rather than a direct re-estimation of the 2016--2024 effect. It constructs cohort-by-relative-period effects before averaging them, reducing sensitivity to heterogeneous effects and to the implicit weighting of the pooled stacked regression. Restricting controls to states with no previous female nomination tests whether earlier nominations have persistent effects that contaminate the comparison group. County-by-cohort fixed effects absorb permanent differences across counties within each event block and test whether local geographic composition accounts for the estimates. Dropping races in which both major-party nominees are women removes cases where, because party vote shares are nearly complementary, the estimated association for one party might partly capture the gender of the opposing nominee. Finally, the experience-adjusted Republican specification asks whether observable differences in candidate qualifications explain the result. 

The upper rows produce a consistent partisan contrast. For Republicans, the full-panel interaction-weighted estimate is \(-4.94\) percentage points (\(SE=2.09\)); the recent estimates are \(-6.06\) when previously treated controls are excluded, \(-6.53\) with county-by-cohort fixed effects, \(-7.40\) after excluding dual-female races, and \(-6.32\) after adjusting for candidate experience. Their confidence intervals exclude zero. For Democrats, the corresponding estimates range from \(-0.13\) to \(+1.76\) percentage points, and every confidence interval includes zero. The figure therefore shows that both the Republican penalty and the absence of a comparable Democratic differential are stable across these alternative specifications.

The final rows assess the influence and weighting of individual cohorts. The leave-one-cohort-out range reports the estimates obtained by removing each nomination switch in turn. Its entirely negative Republican range, from \(-7.31\) to \(-4.97\), shows that no single cohort produces the result. The equal-cohort estimate gives each switch the same weight instead of using the pooled regression's implicit weights. Its value of \(-6.83\) indicates that the Republican estimate is not generated by unusually influential cohorts. For Democrats, the leave-one-out estimates remain centered near zero, while the equal-weight estimate of \(+2.79\) is presented without an uncertainty interval and should not be interpreted as evidence of a Democratic bonus. Randomization inference, reported in the text, reassigns the treated state within each cohort while preserving the block structure. The resulting \(p\)-values are \(0.004\) for Republicans and \(0.986\) for Democrats. This supports the conventional inference conditional on the block design, but it does not resolve the selection and parallel-trends limitations discussed above.

\begin{center}
\textit{[Figure \ref{fig:robustness} approximately here]}
\end{center}

\textit{Identifying variation and cohort checks.} Figure \ref{fig:cohorts} plots the cohort-specific estimates behind the pooled recent stacks, one regression per switch cohort on its own two-period block. The Democratic estimates run in both directions, ten negative and nine positive, with a median of \(-0.67\). All thirteen Republican estimates are negative, with a median of \(-5.39\); excluding the two just-identified cohorts that rest on a single control state, eleven of eleven are negative with a median of \(-3.50\). The period-specific TWFE estimates in Table \ref{tab:temporal} are, by contrast, descriptive benchmarks. A within-window fixed-effects estimate can use only states with at least two elections inside the window and a nominee-gender switch between them, and that count is small. The 2020--2024 Republican estimate rests on two such states and the 2010--2015 Democratic estimate on eleven. 

\begin{figure}[H]
\centering
\includegraphics[width=\textwidth]{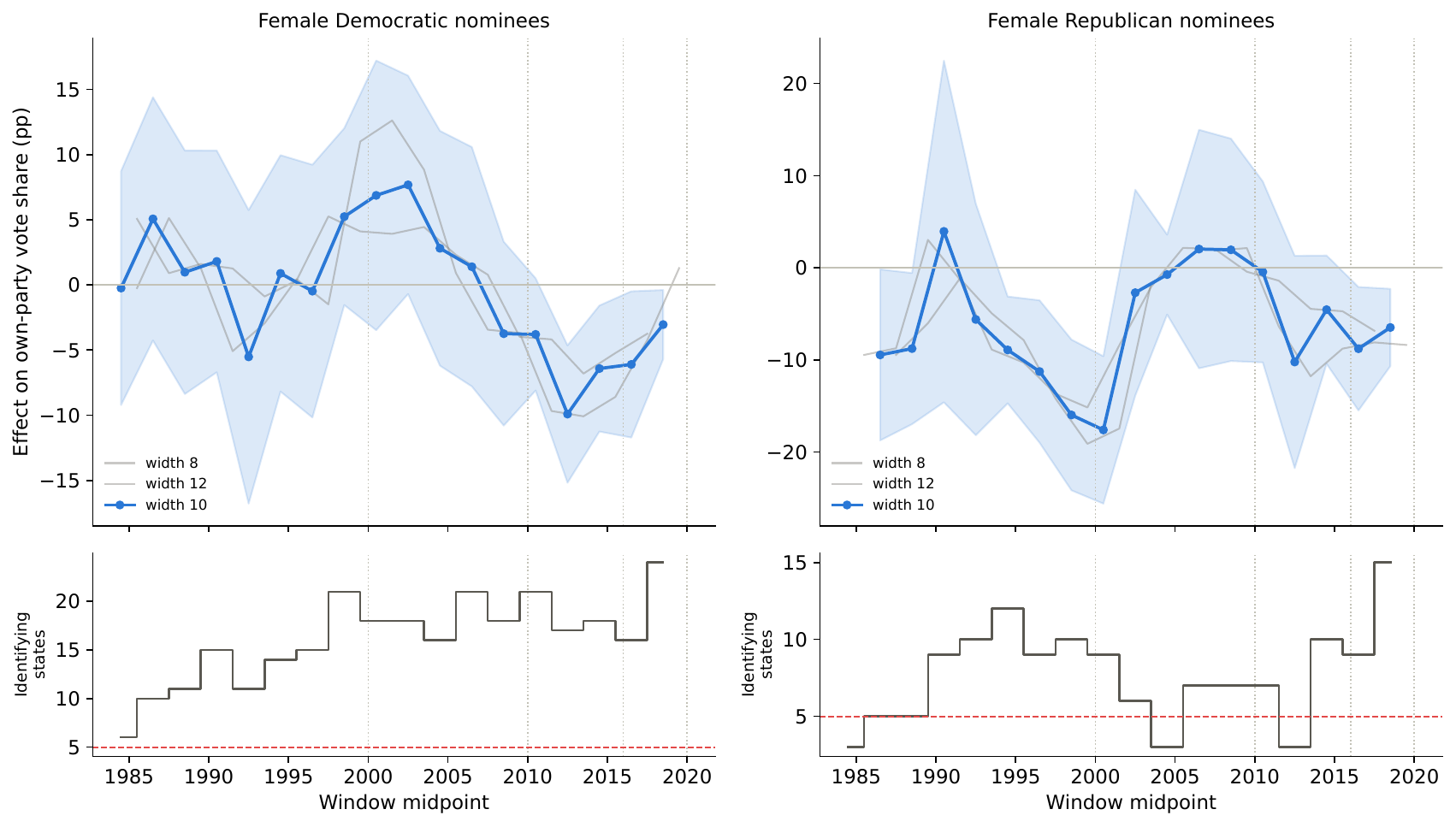}
\caption{Rolling-Window TWFE Estimates and the Variation Behind Them}\label{fig:rolling}
\end{figure}

Figure \ref{fig:rolling} makes the point by sliding a window of fixed width across the panel and plotting, beneath each estimate, the number of states that identify it. The Democratic curve dips around 2010--2015. The six-year window itself gives \(-9.96\) (\(SE = 3.43\), eleven identifying states), and the minimum of the ten-year rolling curve is \(-9.91\), for 2008--2017. That minimum is the product of a search over seventeen overlapping windows. A test that holds the effect fixed at its full-panel value, randomizes the timing of female nominations within each state and recomputes the entire ten-year rolling curve 2,000 times finds a minimum at least as deep as \(-9.91\) in 15.7 percent of draws (\(p = 0.157\)). The stacked design is less exposed to the thin-window problem because each cohort uses the previous election as its baseline and retains every estimable \(0 \rightarrow 1\) switch with clean controls.

For Republican nominees, fourteen of the seventeen estimable ten-year windows produce negative estimates, including every window centered after 2010. The magnitude nevertheless varies substantially over time, and the confidence intervals widen where few states identify the estimate. The Republican panel therefore supports a predominantly negative association across periods, while cautioning against interpreting 2016 as a sharply identified structural break.

\begin{center}
\textit{[Figures \ref{fig:cohorts} and \ref{fig:rolling} approximately here]}
\end{center}

\begin{figure}[H]
\centering
\includegraphics[width=\textwidth]{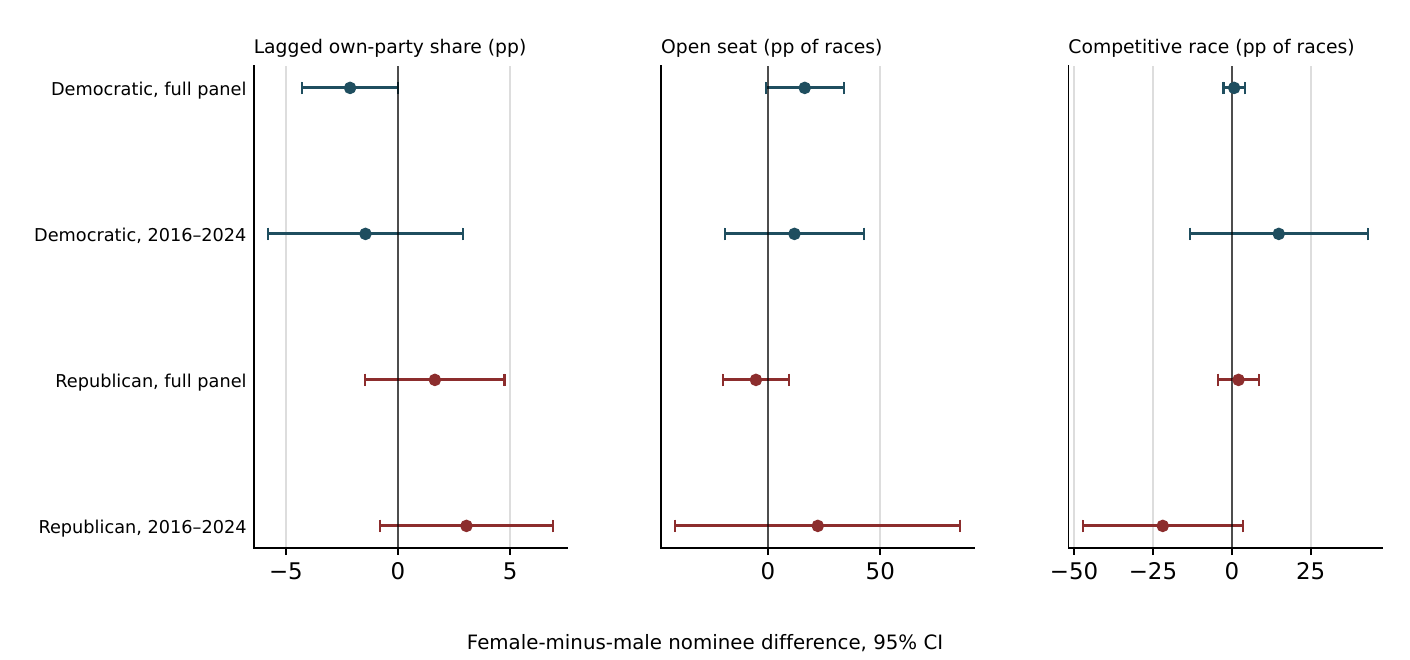}
\caption{Selection on Observables: Electoral Context by Nominee Gender}\label{fig:selection}
\end{figure}

\textit{Selection and candidate quality.} The most direct observable-selection test asks whether female nominees enter worse races. Figure \ref{fig:selection} regresses three pre-determined features of the race --- the own-party lagged vote share, an open-seat indicator and an indicator for a competitive prior result --- on the female-nominee indicator, with state and year fixed effects, on the full panel and on the 2016--2024 state-years the stacked design is built from. For Democratic nominees selection on observables is present but points in competing directions. Female Democratic nominees follow lower lagged Democratic shares (\(-2.1\) pp, \(p < 0.10\)) but are more likely to run in open seats (\(+16.4\) pp, \(p < 0.10\)). The Republican pattern gives little support to the benign interpretation of the headline result. Female Republican nominees do not run after lower lagged Republican shares. The coefficient is positive rather than negative (\(+1.6\) pp and not significant), open seats and prior competitiveness do not differ, and in the 2016--2024 state-years none of the three differences reaches five percent. This does not prove selection away, but it weakens the specific claim that the Republican estimate merely reflects women being nominated in races the party was already poised to lose. The \citet{oster2019} proportional-selection bound points the same way. An unobserved confounder proportional to the included controls would need \(\delta = -8.18\) to move the full-panel Republican estimate to zero, that is, to be several times stronger than the observables and to act against the observed selection pattern.

Candidate quality is a more serious concern. Republican women in the thirteen switch blocks have less measured experience than men (6.4 versus 11.0 mean years in prior elected office, and 23\% versus 57\% with statewide executive experience). Adjusting for those observables, coded under the same rule for male and female nominees, changes the incumbency-adjusted county estimate only modestly, from \(-6.50\) to \(-6.32\) pp (\(SE = 2.03\)), so measured experience does not explain the differential.

\begin{center}
\textit{[Figure \ref{fig:selection} approximately here]}
\end{center}

\begin{figure}[H]
\centering
\includegraphics[width=\textwidth]{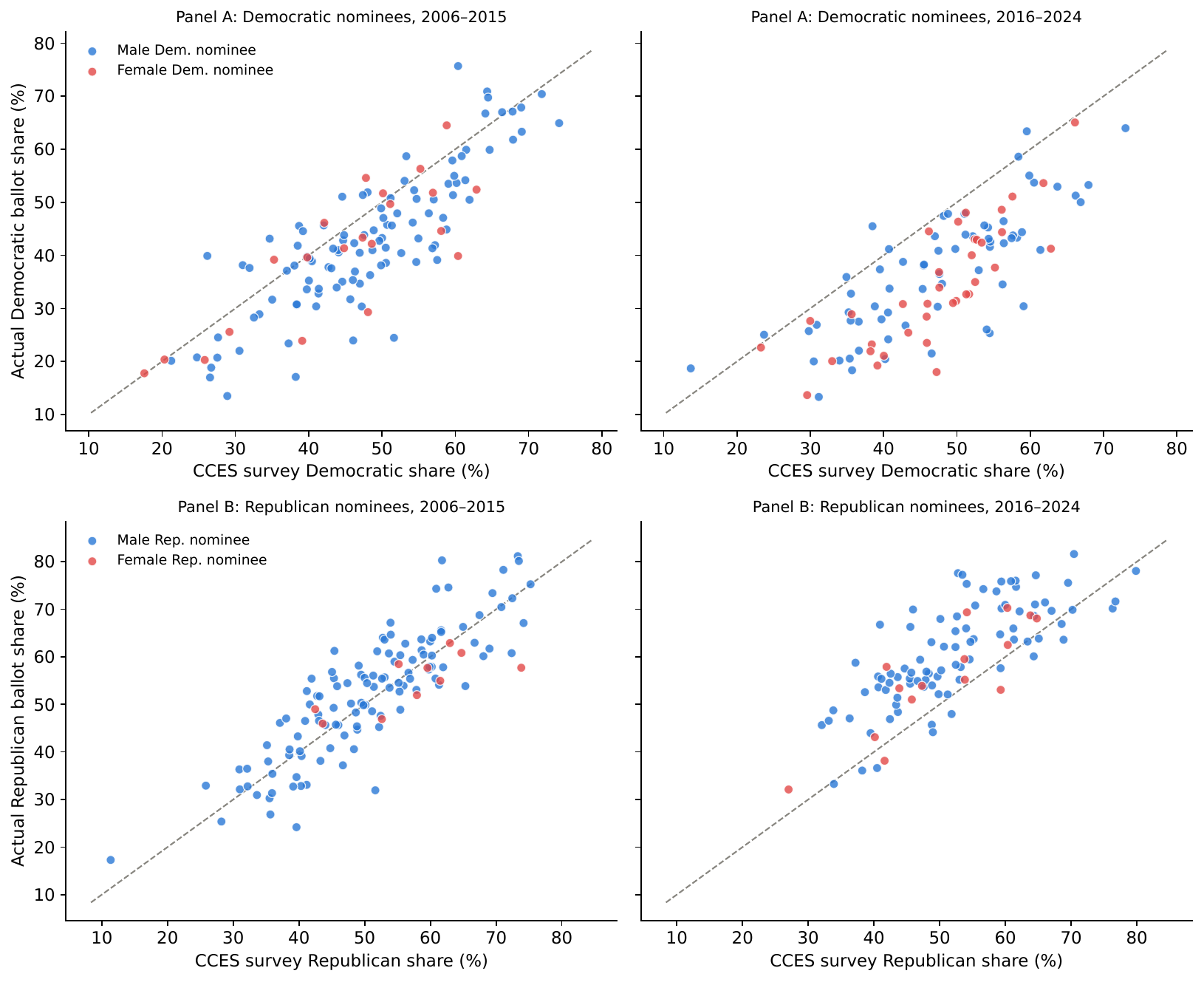}
\caption{Survey-Reported versus Actual Gubernatorial Vote Share, by Party and Era}\label{fig:cces_gap}
\end{figure}

\textit{Survey-ballot diagnostics.} Figure \ref{fig:cces_gap} plots, for 230 state-years from 2006 to 2024, the party's CCES-reported gubernatorial support against its realised ballot share. A point below the 45-degree line is a race in which the ballots delivered less than the survey reported. In the regression counterpart, Republican female nominees receive more reported survey support than their ballot shares would predict (\(+2.4\) pp with state and year fixed effects, significant at five percent, falling to \(+1.9\) pp after partisan-lean and education controls), while the comparable Democratic gap is small and insignificant. The party with the persistent vote-share differential is therefore also the party with the measurable survey-ballot wedge. The pattern is compatible with demand-side discrimination, but it is also compatible with differential reporting or composition. An individual-level test on respondents who report both a gubernatorial and a US House vote on the same ballot finds no corresponding own-party defection when the Republican nominee is a woman, and the CCES extract used here lacks validated turnout. The survey evidence therefore motivates the interpretation problem but does not solve it.

Voters' gender available in surveys' responses provides a natural avenue for further research. In mixed-gender races, male and female voters can be compared within the same state-year and party-identification category. Preliminary CCES estimates suggest some party-specific differences in reported gubernatorial vote choice, but these differences are not robust when gubernatorial choice is differenced from the same respondent's US House vote. Because the available extract also lacks House-candidate gender and validated turnout, and contains relatively few female-Republican races, the survey data do not support the hypothesis that voter--candidate gender matching explains the aggregate pattern. 

\begin{center}
\textit{[Figure \ref{fig:cces_gap} approximately here]}
\end{center}

\textit{Power and the interpretation of nulls.} Power matters mainly for interpreting the nulls. Evaluated at each specification's estimated standard error, rather than as observed power, which is only a transformation of the \(p\)-value \citep{hoenig2001}, the full-panel Democratic specification has 97\% power against a five-point penalty and essentially full power against a seven-point penalty, but only 63\% power against a three-point effect. The stacked Democratic null is similar. The paper can therefore rule out a Democratic penalty on the scale of the Republican estimate, but not smaller differentials. The Republican pre-trend diagnostic is also only moderately powered. As mentioned, it rests on 13 switch events, and one pre-period coefficient is already negative and statistically significant. 

\section{Conclusion}\label{sec:conclusion}

The central finding of the paper is an asymmetric female-nomination performance
differential across the two major parties in US gubernatorial elections. For recent Republican nomination transitions, the stacked estimate
is \(-6.5\) percentage points on the county panel and
\(-3.5\) points on the state-year panel, where it is imprecisely estimated. The
interaction-weighted estimate
is \(-4.9\), and every cohort-specific estimate is negative. By
contrast, the Democratic full-panel and recent-cohort stacked estimates
are indistinguishable from zero. The Democratic trough around
2010--2015 is not statistically unusual after correcting for the search
across rolling windows and is therefore descriptive rather than a
separate finding.

Rebuilding the panel data and extending the coverage proved important. Reconstructing the county panel from primary sources moved one
headline result by more than the effect being measured, because the
coding defects were not randomly distributed but concentrated in the
races with a female nominee and in the most recent decade, which is to
say in exactly the cells that identify the estimates. Extending the time and spatial coverage allowed to capture specific periods' and parties' effects that proved difficult to find in previous research.

Observable candidate experience does not explain the Republican differential.
When prior elected experience is coded under the same rule for male and female
nominees in the stacked blocks, the incumbency-adjusted county estimate changes
by less than one-quarter of a point and remains near \(-6\). The survey-ballot wedge
points in a compatible direction, but it is supporting evidence rather
than independent identification because it cannot separate voter
defection, differential reporting and participation differences. Survey data
for respondents who cast votes in both gubernatorial and US House elections
also show that loyalty does not shift between offices when the Republican
nominee is a woman.

The design identifies the reduced-form consequence of a party
nominating a woman. It does not identify voter
bias holding all other candidate attributes fixed. Strategic nomination
in the transition cycle remains possible, and the parallel-trends assumption is
not satisfied outright. The most defensible
conclusion is therefore narrower than a
universal voter-penalty claim. Recent Republican transitions to female
gubernatorial nominees are followed by a sizable and robust decline in
party vote share relative to clean comparison states, while no
corresponding average differential is detected for Democratic nominees.
Distinguishing voter response from party and candidate selection will
require voter-level data or a design with plausibly exogenous variation
in nominee gender.

These findings contribute to the literature by showing how average null effects
can conceal heterogeneity across parties, periods and types of office. Much of
the existing evidence comes from legislative elections or experiments that
estimate an average response to candidate gender, while work on executive
office suggests that gendered leadership expectations may be more consequential
\citep{eagly2002, anzia2022}. The gubernatorial setting combines an executive
office with repeated elections across states and over 45 years. It therefore
makes it possible to document that the female-nomination differential is neither
stable over time nor symmetric across parties. The recent Republican pattern is
consistent with research on gendered party images and candidate-gender cues
\citep{winter2010, koch2002, king1997}, but the paper does not require any one of
those mechanisms to account for the reduced-form result.

The paper also contributes to the political-selection literature by placing the
nomination decision, rather than candidate gender in isolation, at the center of
the empirical object. Existing work emphasizes that women enter different races,
face different recruitment constraints and may be positively selected on
quality \citep{fulton2009, fulton2014, sanbonmatsu2006}. By combining a rebuilt
and auditable election panel with recurrent-transition estimates, aggregation
checks and candidate-experience diagnostics, this study measures the electoral
performance associated with the nominees parties actually select while making
clear what election returns alone cannot identify. The resulting contribution is
therefore a party- and period-specific benchmark for theories of voter demand and
candidate supply, rather than a universal estimate of voter bias.

\newpage

\phantomsection\label{references}
\addcontentsline{toc}{section}{References}

\begingroup
\setstretch{1.0}
\setlength{\bibsep}{2pt plus 1pt}
\bibliography{references.bib}
\endgroup

\newpage

\section*{Tables}\label{tables}
\addcontentsline{toc}{section}{Tables}


\begin{table}[H]
\centering
\caption{Summary Statistics: Full County-Level Panel 1980--2024}
\TL{summary}
\begin{adjustbox}{max width=\textwidth}
\begin{threeparttable}
\begin{tabular}{lcccccc}
\toprule
Variable & Mean & Std.\ Dev. & Min & Median & Max & N \\
\midrule
Democratic vote share, all ballots \textit{(Panel A outcome)} & 42.968 & 16.037 & 1.08 & 42.68 & 92.18 & 35,076 \\
Republican vote share, all ballots \textit{(Panel B outcome)} & 53.549 & 15.622 & 3.24 & 53.95 & 97.09 & 35,134 \\
Democratic vote share (two-party) & 44.419 & 16.128 & 1.10 & 44.11 & 94.65 & 35,134 \\
Republican vote share (two-party) & 55.581 & 16.128 & 5.35 & 55.89 & 98.90 & 35,134 \\
Lagged Democratic share (two-party) & 45.515 & 15.701 & 1.10 & 45.18 & 94.65 & 31,929 \\
Lagged Republican share (all ballots) & 52.487 & 15.220 & 3.24 & 52.96 & 96.77 & 31,929 \\
Democratic incumbent & 0.249 & 0.432 & 0.00 & 0.00 & 1.00 & 35,135 \\
Republican incumbent & 0.323 & 0.468 & 0.00 & 0.00 & 1.00 & 35,135 \\
Open seat & 0.428 & 0.495 & 0.00 & 0.00 & 1.00 & 35,135 \\
Female Democratic nominee & 0.169 & 0.375 & 0.00 & 0.00 & 1.00 & 35,135 \\
Female Republican nominee & 0.059 & 0.236 & 0.00 & 0.00 & 1.00 & 35,135 \\
\bottomrule
\end{tabular}
\begin{tablenotes}[flushleft]\footnotesize
\item 35,135 county-elections across 578 state-year races in 49 states, 1980--2024. 103 state-years have a female Democratic nominee and 48 a female Republican nominee; 10 races have both, so 141 distinct races carry a female major-party nominee on at least one ticket. The dependent variables are the all-ballot shares in the first two rows. The two lagged controls are built from different bases --- the Democratic lag from the two-party share, the Republican lag from the all-ballot share --- which is inherited from the source panel of \citet{algara2021}. Rebuilding either on the other base leaves both headline estimates in place: Panel A moves from $+0.32$ to $+0.18$ and Panel B from $-4.07$ to $-3.53$. 31,929 observations carry a lagged share and enter the specifications that control for it. Alaska is excluded throughout (it reports by borough); the District of Columbia has no gubernatorial office.
\end{tablenotes}
\end{threeparttable}
\end{adjustbox}
\end{table}

\newpage
\begin{table}[H]
\centering
\caption{TWFE Estimates: Control Specification Sensitivity, Full Panel 1980--2024}
\TL{main}
\begin{adjustbox}{max width=\textwidth}
\begin{threeparttable}
\begin{tabular}{lcccc}
\toprule
& (1) & (2) & (3) & (4) \\
\midrule
\multicolumn{5}{l}{\textbf{Panel A: Female Democratic Nominee}} \\
Female Democratic Nominee & $-2.030$ & $-0.287$ & $-0.883$ & $+0.317$ \\
 & $(1.698)$ & $(1.579)$ & $(1.256)$ & $(1.307)$ \\
Lagged Dem.\ share & --- & $+0.567^{***}$ & --- & $+0.529^{***}$ \\
 &  & $(0.040)$ &  & $(0.044)$ \\
Dem.\ incumbent & --- & --- & $+9.846^{***}$ & $+6.031^{***}$ \\
 &  &  & $(0.844)$ & $(1.070)$ \\
\addlinespace
N & 35,076 & 31,870 & 35,076 & 31,870 \\
$R^2$ & 0.294 & 0.507 & 0.352 & 0.527 \\
\addlinespace
\multicolumn{5}{l}{\textbf{Panel B: Female Republican Nominee}} \\
Female Republican Nominee & $-3.711^{*}$ & $-4.237^{*}$ & $-3.522^{*}$ & $-4.068^{*}$ \\
 & $(1.830)$ & $(1.681)$ & $(1.659)$ & $(1.536)$ \\
Lagged Rep.\ share & --- & $+0.588^{***}$ & --- & $+0.556^{***}$ \\
 &  & $(0.027)$ &  & $(0.033)$ \\
Rep.\ incumbent & --- & --- & $+8.463^{***}$ & $+5.198^{***}$ \\
 &  &  & $(1.308)$ & $(1.244)$ \\
\addlinespace
N & 35,134 & 31,928 & 35,134 & 31,928 \\
$R^2$ & 0.287 & 0.523 & 0.336 & 0.541 \\
\bottomrule
\end{tabular}
\begin{tablenotes}[flushleft]\footnotesize
\item State and year fixed effects throughout. Panel A regresses the Democratic share of all ballots on a female-Democratic-nominee indicator; Panel B the Republican share of all ballots on a female-Republican-nominee indicator. Each panel controls for its own party's incumbent only. Lagged shares are within-county, one election back, and are missing where the gap exceeds six years, which is why columns (2) and (4) use a smaller sample. Standard errors clustered by state in parentheses. $^{\dagger}$ $p<0.10$, $^{*}$ $p<0.05$, $^{**}$ $p<0.01$, $^{***}$ $p<0.001$.
\end{tablenotes}
\end{threeparttable}
\end{adjustbox}
\end{table}

\newpage
\begin{table}[H]
\centering
\caption{Temporal Heterogeneity: TWFE Estimates Across Periods}
\TL{temporal}
\begin{adjustbox}{max width=\textwidth}
\begin{threeparttable}
\begin{tabular}{llcccc}
\toprule
Period & Party & County estimate & State-year estimate & Treated state-years & $N_{SY}$ \\
\midrule
Full period (1980--2024) & Democratic & $+0.32$ & $-1.31$ & 103 & 527 \\
 & & $(1.31)$ & $(1.33)$ & & \\
Full period (1980--2024) & Republican & $-4.07^{*}$ & $-1.59$ & 48 & 527 \\
 & & $(1.54)$ & $(1.35)$ & & \\
\addlinespace
Pre-2016 (1980--2015) & Democratic & $+0.91$ & $-0.90$ & 67 & 414 \\
 & & $(1.82)$ & $(1.49)$ & & \\
Pre-2016 (1980--2015) & Republican & $-4.67^{\dagger}$ & $-2.01$ & 32 & 414 \\
 & & $(2.43)$ & $(1.64)$ & & \\
\addlinespace
Recent (2016--2024) & Democratic & $+1.91$ & $+2.68$ & 36 & 113 \\
 & & $(2.37)$ & $(3.14)$ & & \\
Recent (2016--2024) & Republican & $-7.86^{*}$ & $-1.69$ & 16 & 113 \\
 & & $(3.02)$ & $(2.35)$ & & \\
\bottomrule
\end{tabular}
\begin{tablenotes}[flushleft]\footnotesize
\item Each row is a separate TWFE regression using the same specification as column (4) of Table \ref{tab:main}: own-party lagged share, own-party incumbent, and state and year fixed effects. The periods match the cohort sets in Table \ref{tab:stacked_windows}. County estimates use the county panel; state-year estimates collapse the panel to the assignment level before estimation and use the small-cluster $t$ correction, which evaluates the state-clustered $t$ statistic against a $t$ distribution with $G-1$ degrees of freedom. $N_{SY}$ counts state-year observations in the collapsed panel. Standard errors clustered by state in parentheses. $^{\dagger}$ $p<0.10$, $^{*}$ $p<0.05$, $^{**}$ $p<0.01$, $^{***}$ $p<0.001$.
\end{tablenotes}
\end{threeparttable}
\end{adjustbox}
\end{table}

\newpage
\begin{table}[H]
\centering
\caption{Stacked DiD Estimates Across Cohort Sets}
\TL{stacked_windows}
\begin{adjustbox}{max width=\textwidth}
\begin{threeparttable}
\begin{tabular}{llcccc}
\toprule
Cohort set & Party & County estimate & State-year estimate & Cohorts & $N_{SY}$ \\
\midrule
Full period (1980--2024) & Democratic & $+1.35$ & $+0.91$ & 68 & 2,400 \\
 & & $(1.39)$ & $(2.29)$ & & \\
Full period (1980--2024) & Republican & $-5.26^{*}$ & $-3.32$ & 35 & 1,560 \\
 & & $(2.24)$ & $(2.90)$ & & \\
\addlinespace
Pre-2016 (1980--2015) & Democratic & $+2.20$ & $+1.39$ & 49 & 1,804 \\
 & & $(2.03)$ & $(2.83)$ & & \\
Pre-2016 (1980--2015) & Republican & $-4.28$ & $-3.22$ & 22 & 992 \\
 & & $(3.50)$ & $(4.03)$ & & \\
\addlinespace
Recent (2016--2024) & Democratic & $-0.03$ & $+0.82$ & 19 & 596 \\
 & & $(1.62)$ & $(3.53)$ & & \\
Recent (2016--2024) & Republican & $-6.50^{**}$ & $-3.47$ & 13 & 568 \\
 & & $(1.90)$ & $(2.62)$ & & \\
\bottomrule
\end{tabular}
\begin{tablenotes}[flushleft]\footnotesize
\item Each row uses the same two-period stacked construction around $0\to1$ nomination switches, with controls untreated in both years of the cohort block and state$\times$cohort and year$\times$cohort fixed effects. All estimates include the own-party incumbency control used in Table \ref{tab:temporal}; no lagged share enters because, in a two-period stack, the pre-period outcome is the baseline. The full-period cohort set includes every estimable switch from 1980 to 2024; the pre-2016 and recent rows split those cohorts at 2016. County estimates use the county panel; state-year estimates collapse the identical cohort blocks to the assignment level before estimation and use the small-cluster $t$ correction, which evaluates the state-clustered $t$ statistic against a $t$ distribution with $G-1$ degrees of freedom. $N_{SY}$ counts state-year-cohort observations in the collapsed stack. Standard errors clustered by state in parentheses. $^{\dagger}$ $p<0.10$, $^{*}$ $p<0.05$, $^{**}$ $p<0.01$, $^{***}$ $p<0.001$.
\end{tablenotes}
\end{threeparttable}
\end{adjustbox}
\end{table}

\newpage
\begin{table}[H]
\centering
\caption{The Party Asymmetry Estimated Directly}
\TL{asymmetry}
\begin{adjustbox}{max width=\textwidth}
\begin{threeparttable}
\begin{tabular}{llcccc}
\toprule
Cohort set & Level & Democratic effect & Republican $-$ Democratic & Cohorts & $N$ \\
\midrule
Full period (1980--2024) & County & $+1.31$ & $-6.50^{*}$ & 103 & 246,388 \\
 & & $(1.40)$ & $(2.81)$ & & \\
Full period (1980--2024) & State-year & $+1.20$ & $-4.89$ & 103 & 3,946 \\
 & & $(2.42)$ & $(3.31)$ & & \\
\addlinespace
Pre-2016 (1980--2015) & County & $+2.36$ & $-6.94$ & 71 & 175,481 \\
 & & $(1.96)$ & $(4.38)$ & & \\
Pre-2016 (1980--2015) & State-year & $+1.78$ & $-5.46$ & 71 & 2,796 \\
 & & $(2.89)$ & $(4.99)$ & & \\
\addlinespace
Recent (2016--2024) & County & $+0.02$ & $-6.64^{*}$ & 32 & 70,907 \\
 & & $(1.61)$ & $(2.64)$ & & \\
Recent (2016--2024) & State-year & $+0.95$ & $-5.05$ & 32 & 1,150 \\
 & & $(3.71)$ & $(3.79)$ & & \\
\bottomrule
\end{tabular}
\begin{tablenotes}[flushleft]\footnotesize
\item Both parties' switch blocks are stacked into a single panel for each cohort set, with the own-party vote share as the outcome and the own-party female nomination as the treatment, absorbing state$\times$cohort and year$\times$cohort fixed effects and controlling for own-party incumbency. The first column is the Democratic effect; the second is the interaction with a Republican indicator, which is the difference between the two parties' effects and is the quantity the paper's asymmetry claim is about. Estimating it as a coefficient rather than comparing two separately estimated numbers is what puts a standard error on the asymmetry. The cohort sets and period labels match Table \ref{tab:stacked_windows}. The two levels are the county panel and the state-year collapse of the same blocks. One caveat the design cannot remove is that some races field a woman on both tickets, so the two parties' estimates are not independent in those blocks; in a near-two-candidate race one party's share is also close to the complement of the other's; Figure \ref{fig:robustness} reports the recent-window check that drops such blocks. $N$ counts county-elections across blocks and across both parties' panels, so a county recurs once per block it serves in and again in the other party's stack; it is not a count of distinct races. Standard errors carry the small-cluster $t$ correction ($G-1$ degrees of freedom), which charges the absorbed fixed effects to the degrees of freedom --- on the state-year panel that is the difference between significance and none. Standard errors clustered by state in parentheses. $^{\dagger}$ $p<0.10$, $^{*}$ $p<0.05$, $^{**}$ $p<0.01$, $^{***}$ $p<0.001$.
\end{tablenotes}
\end{threeparttable}
\end{adjustbox}
\end{table}

\newpage
\begin{table}[H]
\centering
\caption{Pre-Trend Diagnostics}
\TL{pretrend}
\begin{adjustbox}{max width=\textwidth}
\begin{threeparttable}
\begin{tabular}{lcc}
\toprule
Diagnostic & FDN & FRN \\
\midrule
Pre-period $k = -3$ & $-1.04$ $(4.15)$ & $-5.41$ $(6.67)$ \\
Pre-period $k = -2$ & $-1.00$ $(2.57)$ & $-4.12^{*}$ $(1.98)$ \\
Switch cycle $k = 0$ & $-0.33$ $(1.49)$ & $-6.59^{**}$ $(1.92)$ \\
\midrule
Joint pre-trend test & $\chi^2(2) = 0.15$, $p = 0.926$ & $\chi^2(2) = 4.34$, $p = 0.114$ \\
Switch events & 23 & 13 \\
$N$ & 63,346 & 69,463 \\
\bottomrule
\end{tabular}
\end{threeparttable}
\end{adjustbox}
\end{table}
\vspace{-0.5em}
{\footnotesize \textit{Notes:} The table reports the event-study diagnostic on the switches the recent stacked estimate is built from --- those from 2016 on --- and in the specification it actually uses: cohort-specific state and year fixed effects, own-party incumbency, and no lagged share, which a two-period stack cannot include because the lag is the pre-period outcome. The omitted reference period is $k=-1$. \textbf{The Republican pre-trend is not zero}: $k=-2$ is $-4.12$ (SE $1.98$), significant at the five percent level, though the joint test over both pre-periods is not ($p = 0.114$). The Democratic side is flat individually and jointly. This is why the Republican estimate is reported as a reduced-form association with an explicit parallel-trends limitation rather than as a fully certified causal effect. Standard errors clustered by state in parentheses. $^{\dagger}$ $p<0.10$, $^{*}$ $p<0.05$, $^{**}$ $p<0.01$, $^{***}$ $p<0.001$.}

\newpage

\section*{Figures}\label{figures}
\addcontentsline{toc}{section}{Figures}

\begin{figure}[H]
\centering
\includegraphics[width=\textwidth]{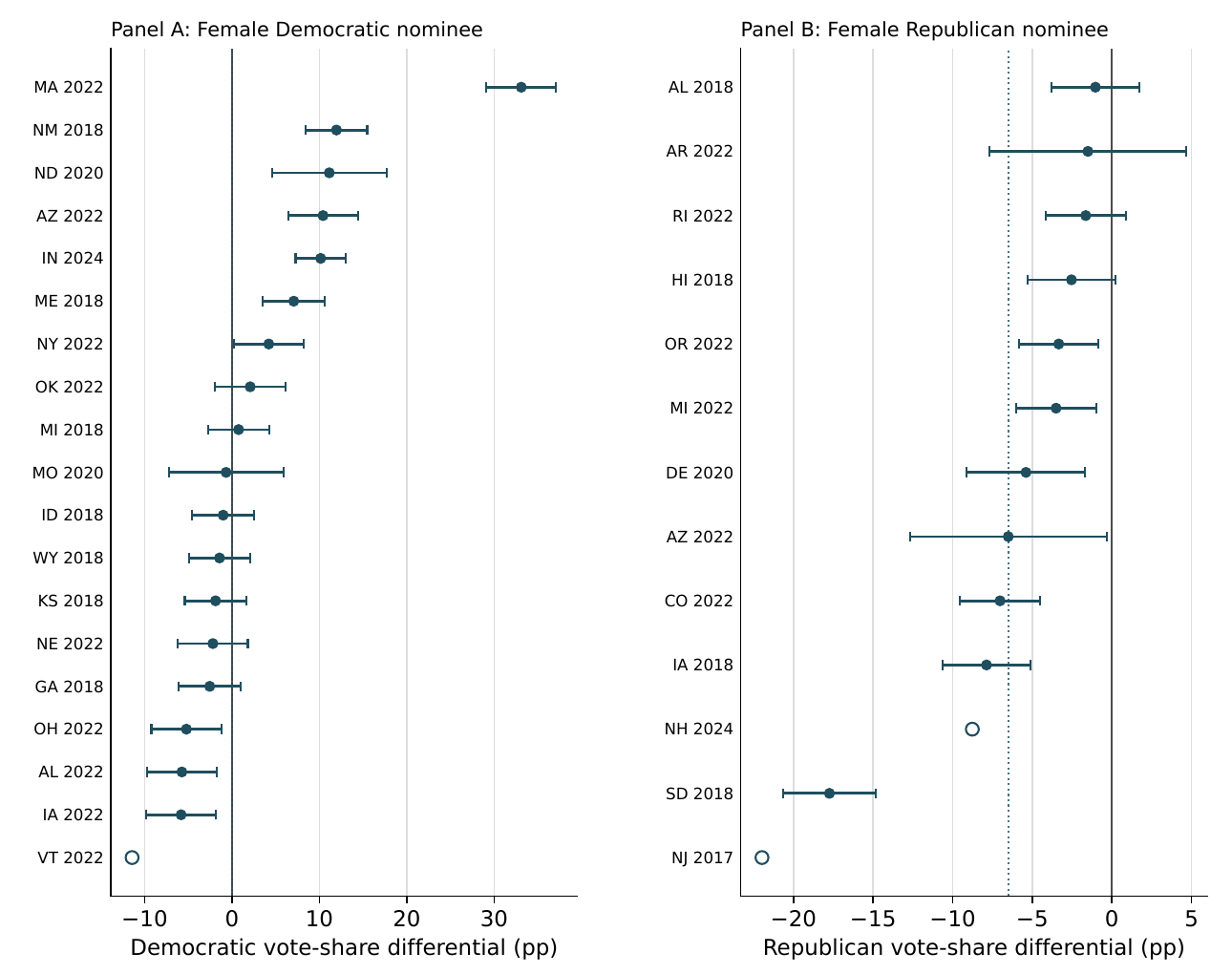}
\caption{Cohort-Specific Stacked Estimates, 2016--2024 Switch Cohorts}\label{fig:cohorts}
\end{figure}

{\small \textit{Notes:} Each panel takes the recent (2016--2024) county-level stacked estimate for one party as its baseline. The interaction-weighted row is the exception: it averages cohort-by-relative-period effects over the full event panel. Whiskers around point estimates are 95 percent confidence intervals based on state-clustered standard errors. ``No-prior controls'' restricts each block's controls to states with no earlier female nomination of that party. ``County x cohort FE'' replaces the state-by-cohort fixed effects with county-by-cohort fixed effects. ``Drop both-female tickets'' excludes blocks in which both major parties nominated women. ``Experience adjusted'' controls for prior elected years and prior statewide executive office coded under the same rule for male and female nominees; the coding exists for the Republican blocks only. The leave-one-cohort-out segment is an influence range, not a confidence interval: it spans the estimates obtained by dropping each switch cohort in turn. The equal-cohort-weight diamond is the unweighted mean of the cohort-specific estimates in Figure \ref{fig:cohorts} and has no confidence interval. The dotted vertical line marks the county-level baseline in Table \ref{tab:stacked_windows}.}

\newpage

{\small \textit{Notes:} One regression per switch cohort on its own two-period block, controlling for the own-party incumbent, with state and year fixed effects within the block; the outcome is the party's share of all ballots. Cohorts are labelled by state and switch year and sorted by estimate. Intervals are 95 percent confidence intervals from state-clustered standard errors. Hollow markers are cohorts with a single clean control state (VT 2022; NH 2024 and NJ 2017), for which the within-block comparison is just-identified and the clustered standard error is degenerate; they are shown without an interval and contribute 48 of 34,747 and 356 of 36,160 observations to the pooled Democratic and Republican stacks, which are estimated jointly rather than as an average of these points. The dotted vertical line marks the pooled county-level estimate of Table \ref{tab:stacked_windows}.}

\newpage

{\small \textit{Notes:} Upper panels re-estimate column (4) of Table \ref{tab:main} on a window of fixed width that slides across the panel, plotted at the window midpoint, for widths of 8, 10 and 12 years; the shaded band is the 95 percent confidence interval for the 10-year window. Lower panels count the states that identify each 10-year estimate, that is, states with two or more elections inside the window and a change in the sex of their nominee between them; the dashed line marks five. The minimum of the ten-year Democratic curve is \(-9.91\), for the 2008--2017 window; the six-year 2010--2015 window discussed in the text gives \(-9.96\) (\(SE = 3.43\), eleven identifying states). Holding the effect fixed at its full-panel value, randomizing the timing of female nominations within each state and recomputing the ten-year curve 2,000 times yields a minimum at least as deep as \(-9.91\) in 15.7 percent of draws.}

\newpage

{\small \textit{Notes:} Each point is the coefficient on the female-nominee indicator of the party shown in a regression of a pre-determined feature of the race on that indicator with state and year fixed effects, with its 95 percent confidence interval from state-clustered standard errors. The lagged share is the own-party share in the state's previous gubernatorial election; a competitive race is one whose own-party lagged share is within five points of 50. Open-seat and competitiveness coefficients are linear-probability estimates expressed in percentage points of races. ``Full panel'' rows use the county panel for 1980--2024 (31,929 observations); ``2016--2024'' rows use the state-year panel for the window the stacked design is built from (114 races, 36 with a female Democratic and 16 with a female Republican nominee).}

\newpage

{\small \textit{Notes:} Each point is a state-year, 230 in all, 2006--2024. The horizontal axis is the party's CCES-reported share among respondents answering the governor question; the vertical axis is its realised share of ballots. The dashed line is the 45-degree line: a point below it is a race in which the ballots delivered less than the survey reported. Panel A plots Democratic nominees and Panel B Republican nominees, each split at 2016. Red markers are races with a female nominee of that party. In the regression counterpart, with state and year fixed effects, the survey-ballot gap is \(+2.4\) pp larger for female Republican nominees (\(SE = 1.1\)), \(+1.9\) pp after partisan-lean and education controls, and not significantly different for female Democratic nominees.}

\end{document}